\documentclass{ws-procs975x65}
\usepackage{multirow}

\def\al{\alpha} 
 
\def\ga{\gamma}

\def\et{\eta}

\def\si{\sigma}

\def\Om{\Omega}

\def\bi{{\mathbf{i}}}

\newcommand{\ben}{\begin{equation}}
\newcommand{\een}{\end{equation}}
\newcommand{\bea}{\begin{eqnarray}}
\newcommand{\eea}{\end{eqnarray}}
\newcommand{\ba}{\begin{array}}
\newcommand{\ea}{\end{array}}
\newcommand{\bit}{\begin{itemize}}
\newcommand{\eit}{\end{itemize}}

\def\math{\mathsurround 0pt}
\def\oversim#1#2{\lower.5pt\vbox{\baselineskip0pt \lineskip-.5pt
        \ialign{$\math#1\hfil##\hfil$\crcr#2\crcr{\scriptstyle\sim}\crcr}}}

\def\pa{\partial}

\def\half{\frac{1}{2}}

\newcommand{\etRG}{{\et_{\rm RG}}}

\newcommand{\nuRG}{{\nu_{\rm RG}}}
\newcommand{\alRG}{{\al_{\rm RG}}}
\newcommand{\siRG}{{\si_{\rm RG}}}

\begin{document}

\title{\uppercase{The asymptotic safety scenario and scalar field inflation}}

\author{\uppercase{Christoph Rahmede$^*$}}

\address{Karlsruhe Institute of Technology,  Institute for Theoretical Physics,  Wolfgang-Gaede-Str. 1, 76128 Karlsruhe, Germany
\\
Institut f\"ur Physik, TU Dortmund, Otto-Hahn-Str. 4, 44221 
Dortmund, Germany\\
$^*$E-mail: christoph.rahmede@kit.edu}

\begin{abstract}
We study quantum gravity corrections to early universe cosmology as resulting within the asymptotic safety scenario. 
We analyse if it is possible to obtain
accelerated expansion in the regime of the renormalisation group fixed point in a theory with Einstein-Hilbert gravity 
and a scalar field. 
We show how this phase 
impacts cosmological perturbations observed in the cosmic microwave background.
\end{abstract}


\bodymatter
\bigskip

A viable quantum gravity scenario based on standard quantum field theory methods can be obtained if the Renormalisation Group
 (RG) flow of gravity 
is controled by a fixed point (FP) at very high energies. If such a FP has only a finite number of attractive directions,
it can only be reached from a finite dimensional subspace of coupling space, the so-called UV critical surface to which the RG 
flow has to be confined. In that case 
there exist only a finite number of free parameters
and the theory is predictive. 
To realise a theory of quantum gravity along this idea has been proposed by S. Weinberg as the asymptotic safety scenario\cite{Weinberg:1980gg}. 

In the recent years, increasing evidence for the existence of a suitable RG FP has been found with functional RG 
techniques in gravity\cite{Reuter:1996cp,Niedermaier:2006wt,Percacci:2007sz,Reuter:2012id,Litim:2008tt,Codello:2008vh,Falls:2013bv}
and gravity with scalar matter
\cite{Narain:2009fy,Narain:2009gb}. 
A natural question  in that context is if the presence of a RG FP can lead to distinctive predictions for early
universe cosmology with potential 
observable signals e.g. in the cosmic microwave background (CMB). 

In the asymptotic safety approach, naturally all interaction operators consistent with the underlying symmetry principles come 
into play at high energies, although their couplings will not be independent. Therefore, it should be possible to create 
inflationary  expansion in the early universe with the help of RG corrections. An efficient expansion mechanism could take place 
as early as during the RG FP regime when RG effects are most significant, or also at later stages if appropriate values for
the couplings are reached along the RG flow.  Inflation could result either from purely gravitational interaction terms with some resemblance to Starobinsky 
inflation, or from some matter ingredient like the inflaton.
The first possibility has been considered e. g. in Refs.~\refcite{Bonanno:2001xi,Bonanno:2010bt,Weinberg:2009wa,Hindmarsh:2012rc},
%Recent calculations of 
%renormalisation group flows of purely gravitational actions of the polynomial $f(R)$-type \cite{Falls:2012} indicate that
%de Sitter-solutions in the RG FP regime are not obtained beyond some limited order ($R^{11}$) and therefore might not be 
%obtained in general. 
%In that case, de Sitter stages might still be reached along the RG flow, but we
%take this as a strong motivation to use some
%matter component like an inflaton field  to obtain inflationary expansion in the RG FP regime. 
whereas in Refs.~\refcite{Hindmarsh:2011hx,Contillo:2011ag} the second approach was taken in order to be as close as possible
to standard model cosmology.

For such models, based on the assumption that FRW-cosmology is still a good description 
in the RG FP regime, there exist FPs of the dynamical equations (also classically) which allow for accelerated expansion,
and we can calculate the resulting spectrum of cosmological perturbations. These perturbations should remain small. If they turn out
to be too large, on the one hand spacetime in the FP regime would show significant deviations from the FRW-assumptions 
of homogeneity and isotropy,
on the other hand they could not be responsible for the perturbations observed in the CMB spectrum
and would have to be washed out by later stages of inflation not obtained during the FP regime.  

For a simple model of how inflation could be created during the RG FP regime, we consider the standard model of 
cosmology based on the Einstein-Hilbert action, a minimally coupled scalar field $\phi$ with potential $V$
\cite{Hindmarsh:2011hx} and some ideal fluid component. The classical equations of motion for this model in terms of the dimensionless
variables
$x = \frac{\kappa\dot{\phi}}{\sqrt{6}H}$, $y=\frac{\kappa\sqrt{V}}{\sqrt{3}H}$,  $z = \frac{V'}{\kappa V}$, 
$\Omega_i=\frac{\kappa^2\rho_i}{3 H^2}$, $N=-\ln a$, $\kappa=\sqrt{8\pi G}$,
are
\begin{equation} \label{SMEOMs}
\begin{array}{ccccccc}
\frac{{d}x}{{d}N}&=&
3x(1-x^2)+\sqrt{\frac32}y^2z -\frac{3}{2}x\sum_i\ga_i\Om_i&\ ,&
\frac{{d}z}{{d}N}&=&-\sqrt{6}\,x(\eta-z^2)\ , 
\\
\frac{{d}y}{{d}N}&=&
-\sqrt{\frac32}x yz - 3 x^2 y - \frac{3}{2}y\sum_i\ga_i\Om_i&\ ,&
\frac{d\Om_i}{dN}&=&-3\Om_i\left( 2x^2+\sum_j\ga_j\Om_j - \ga_i \right)\ ,\nonumber
\end{array}
\end{equation}
with the Friedmann constraint
$1 - x^2 - y^2 - \sum_i\Om_i=0$.

RG effects are
assumed to become more and more important at high energies, or equivalently early times. Thus there should exist a mapping
between the RG scale $k$, with which the RG effects increase, and the age of the universe $t$. 
Then RG effects are taken into account
by making the couplings time dependent. The equations of motion for the Hubble parameter $H$ and the scalar field $\phi$ 
then have to be corrected by inserting couplings which 
change with time.
This has to be done however without violating the Bianchi identity which certifies energy-momentum conservation. 
With this requirement we obtain a relation between $k$ and $t$ so that the Bianchi identity remains unchanged. 
Parametrising the RG dependence by the quantities
$\etRG=\frac{\pa\ln G}{\pa\ln k}$, $\nuRG=\frac{\pa \ln V}{\pa \ln k}$, $\siRG = \frac{\pa \ln V'}{\pa \ln k}$,
$\alRG = \half\left[ \etRG + \nuRG - \frac{\partial}{\partial \ln k} \ln \left(-\frac{\etRG}{\nuRG}\right)\right]$
adds then to Eqs. (\ref{SMEOMs}) the correction terms 
\begin{equation}
\begin{array}{ccccccc}
\frac{{d}x}{{d}N}\rvert_{\rm RGcorr}&=&\half x \,\etRG  \frac{d \ln k}{dN}&\ ,
&\frac{{d}y}{{d}N}\rvert_{\rm RGcorr}&=& \half y (\etRG + \nuRG)  \frac{d \ln k}{dN}\ ,\nonumber\\
\frac{{d}z}{{d}N}\rvert_{\rm RGcorr}&=& -z\left(\half \etRG +\nuRG - \siRG\right)  \frac{d \ln k}{dN}&\ ,
&\frac{d\Om_i}{dN}\rvert_{\rm RGcorr}&=&\etRG\Om_i \frac{d \ln k}{dN}\ .
\end{array}
\end{equation}
The Friedmann constraint takes the form
$\etRG(k) + y^2\, \nuRG(k,z) = 0$.
All corrections are proportional to 
\begin{equation}
\frac{d \ln k}{dN}
 =\frac{1}{\alRG}\left[\frac{\siRG}{\nuRG}\sqrt{\frac32}x z + 3 x^2  + \frac{3}{2}\sum_i\ga_i\Om_i  \right]\ .  \label{eq:TeqnRG}
\nonumber
\end{equation}
The FPs of this system of dynamical equations might control the very early or very
late stages of cosmology and might provide accelerated expansion. One distinguishes two cases: \\
a) For some $N$ 
one has $d\ln k/d N=0$, and the evolution of $k$ with $N$ comes to a halt.
Then the couplings are frozen to some generic value until the dynamical FP 
point is reached for $N\to \infty$.\\
% but the couplings still have generic values.
b) A dynamical FP is reached where $d\ln k/d N={\rm const.}\neq 0$. Then it is reached simultaneously with the
RG FP.\\
Both cases can be realised with explicit types of potentials\cite{Hindmarsh:2011hx}. 
Domination by the potential term of the scalar field ensuing 
accelerated expansion can be obtained in several of these cases.

In the particular case of a quartic potential $V=\lambda_0+\lambda_2\phi^2+\lambda_4\phi^4$ and no ideal fluid components, a 
solution where dynamical FP and RG FP are reached simultaneously has been found \cite{Contillo:2011ag} where all quantities scale
with time, $H=\alpha/t$, $\phi=\varphi/t$, $k=\chi/t$, with dimensionless factors fixed by the dynamical equations. 
The cosmological perturbations created during the RG FP regime can be calculated for this case similarly to 
the standard procedure after checking that the same quantum mode functions can be used and that the perturbations are conserved after
horizon exit. Then one obtains for the curvature and tensor perturbations for wave number $p$ and conformal time $\tau$
\begin{equation}
{\cal P}_{\cal R }(p)=\frac{1}{\pi}\frac{{\tilde G}\alpha^3}{\chi^2}\left(1-\frac{1}{\alpha}\right)^2\ ; \ 
{\cal P}_{t }(p) \simeq \frac{4\tilde G}{\pi} \frac{\alpha^2}{\chi^2}\left(-p\tau\right)^{n_T}
\end{equation}
with scale invariant curvature perturbation and the tensor tilt $n_T=-2/(\alpha-1)$. Small perturbations can be obtained 
if $\chi^2\gg\tilde G \alpha^3$ with $\tilde G=G k^2$. This implies that arbitrarily small perturbations can be obtained by 
staying close to the line  $\lambda_2=-2\sqrt{\lambda_0\lambda_4}$.
Then, the assumption of FRW cosmology during the RG FP regime can be preserved and
it could be possible that cosmological perturbations are influenced by an inflationary period
taking its origin during this regime.

\section*{Acknowledgements}

C. R. would like to thank E. Mielke for the invitation to this conference and M. Hindmarsh for comments on the manuscript.

\bibliographystyle{ws-procs975x65}

\end{document}